\documentclass[aps,prd,10pt,notitlepage,showpacs,nofootinbib,superscriptaddress,twocolumn,raggedbottom,longbibliography]{revtex4-1}
\usepackage{amsfonts}
\usepackage{amsmath}
\usepackage{bm}
\usepackage{soul}
\usepackage{babel}
\usepackage{booktabs}
\usepackage{dsfont}
\usepackage[export]{adjustbox}% vertical alignment
\usepackage{graphicx}
\usepackage{nicefrac}
\usepackage[dvipsnames]{xcolor}
\usepackage{braket}
\usepackage{enumitem} % for customizing lists
\usepackage{mathtools}
\usepackage{multirow} 
\usepackage{ragged2e} % for caption alignment
\usepackage{float}    % For [H] placement if needed
\usepackage{marginnote}
\usepackage{hyperref}
\hypersetup{
    colorlinks=true,
    urlcolor=blue,
    linkcolor = blue,
    urlcolor  = blue,
    citecolor = blue,
    anchorcolor = blue,
    pdftitle={NN-study},
    pdfauthor={DanielSadasivan},
    pdfsubject={NN}}
\usepackage{cleveref} %use \cref{}
    \crefformat{figure}{Fig.~#2#1#3}
    \crefformat{table}{Tab.~#2#1#3}
    \crefformat{equation}{Eq.~(#2#1#3)}
    \crefformat{section}{Sec.~#2#1#3}
\usepackage{xparse}
\usepackage{etoolbox}

\renewcommand{\Re}{{\rm Re}}
\renewcommand{\Im}{{\rm Im}}
\newcommand{\MeV}{{\rm \,MeV}}

\newcommand{\GeV}{{\rm \,GeV}}
\begin{document}

%%%%%%%%%%%%%%%%%%%%%%%%%%%%%%%%%%%%%%%%%%%%%%%
\title{Deep Neural Network Driven Simulation Based Inference Method for Pole Position Estimation under Model Misspecification}
%%%%%%%%%%%%%%%%%%%%%%%%%%%%%%%%%%%%%%%%%%%%%%%
\author{Daniel Sadasivan}
\email{daniel.sadasivan@avemaria.edu}
\affiliation{Ave Maria University, Ave Maria, FL 34142, USA}

\author{Isaac Cordero}
\email{isaac.cordero@my.avemaria.edu}
\affiliation{Ave Maria University, Ave Maria, FL 34142, USA}
\author{Andrew Graham}
\email{andrew.graham@my.avemaria.edu}
\affiliation{Ave Maria University, Ave Maria, FL 34142, USA}

\author{Cecilia Marsh}
\email{cecilia.marsh@my.avemaria.edu}
\affiliation{Ave Maria University, Ave Maria, FL 34142, USA}

\author{Daniel Kupcho}
\email{daniel.kupcho@my.avemaria.edu}
\affiliation{Ave Maria University, Ave Maria, FL 34142, USA}

\author{Melana Mourad}
\email{melana.mourad@my.avemaria.edu}
\affiliation{Ave Maria University, Ave Maria, FL 34142, USA}

\author{Maxim Mai}
\email{maxim.mai@faculty.unibe.ch}
\affiliation{Albert Einstein Center for Fundamental Physics, Institute for Theoretical Physics, University of Bern, Sidlerstrasse
5, Bern, 3012, Switzerland}
\affiliation{Department of Physics, The George Washington University, 725 21st St, NW, Washington, 20052, District of Columbia, USA}

%%%%%%%%%%%%%%%%%%%%%%%%%%%%%%%%%%%%%%%%%%%%%%%

%%%%%%%%%%%%%%%%%%%%%%%%%%%%%%%%%%%%%%%%%%%%%%%%%%%%%%%
%%%%%%%%%%%%%%%%%%%%%%%%%%%%%%%%%%%%%%%%%%%%%%%%%%%%%%%
\begin{abstract}
The method of Simulation Based Inference is shown to lead to a more accurate resonance parameter estimation than traditional $\chi^2$ minimization in certain cases of model misspecification in a case-study of $\pi\pi$ scattering and the $\rho(770)$-resonance. Models fit to certain data sets using $\chi^2$ minimization can make inaccurate predictions for the pole position of the $\rho(770)$. SBI is shown to make a more robust predictions for the pole positions. This is significant, both as a proof of concept that the SBI method can be used in cases of model misspecification, and because models of $\pi\pi$ scattering are a crucial part to many physical systems of contemporary interest ($a_1(1260)$, $\omega(782)$ etc.).
\end{abstract}
%%%%%%%%%%%%%%%%%%%%%%%%%%%%%%%%%%%%%%%%%%%%%%%%%%%%%%%
%%%%%%%%%%%%%%%%%%%%%%%%%%%%%%%%%%%%%%%%%%%%%%%%%%%%%%%

\maketitle

%%%%%%%%%%%%%%%%%%%%%%%%%%%%%%%%%%%%%%%%%%%%%%%%%%%%%%%
%%%%%%%%%%%%%%%%%%%%%%%%%%%%%%%%%%%%%%%%%%%%%%%%%%%%%%%
\section{Introduction}
A traditional and widely-used method for model parameter determination given data is to find the parameters that minimize $\chi^2$ statistic. Simulation Based Inference (SBI) is an alternative method, developed and used in Refs. \cite{tavare1997inferring, rubin1984bayesianly, diggle1984monte, beaumont2002approximate} and applied to a number of fields including particle physics~\cite{Brehmer:2020cvb}, astrophysics~\cite{Mishra-Sharma:2021nhh,Legin:2021zup}, the Mpemba Effect~\cite{Amorim_2023}, neuroscience~\cite{10.7554/eLife.56261}, climate science and evolutionary biology~\cite{Avecilla}. A pedagogical introduction in this matter can be found in Ref.~\cite{astroautomata2020sbi}. In a nutshell, this method involves using a model to generate synthetic data, also referred to as pseudodata, and then using this pseudodata to construct a probability distribution for the model parameters given the data. The probability distribution is often modeled using deep neural networks. 

Recent research has focused on understanding how SBI performs in cases referred to as misspecification, in which either the data or the model is flawed. Some works have shown challenges with the SBI method~\cite{cannon2022investigatingimpactmodelmisspecification}, and some have used SBI as a method to test for model misspecification~\cite{AnauMontel:2024flo} while others develop versions of SBI that are more robust to model misspecification~\cite{huang2023learningrobuststatisticssimulationbased, wehenkel2024addressingmisspecificationsimulationbasedinference}. The present paper uses a version of SBI that is robust under model misspecification to predict parameters relevant to hadronic physics, first for a toy model and then for the actual data. It should be noted that in an ideal case in which (a) data are normally distributed around a model that is considered and (b) the minimization routine can find the global minimum, the $\chi^2$ method is guaranteed to give more probable fit parameters than the SBI method. However, in certain non-ideal cases, the SBI method might be able to outperform the method of $\chi^2$ minimization. Thus, the aim of this paper is not to replace the standard practice of obtaining parameters through $\chi^2$ minimization, but rather to introduce an alternative that can be useful in certain cases, specifically in the presence of model misspecification.

The specific focus of this work is universal parameters of resonances, unstable particles that can be formed in scattering or decay processes. These parameters are encoded in the pole positions of the transition amplitudes. They are typically obtained by providing a suitable model for the latter respecting principles of S-matrix theory (possibly accommodating further symmetries) and fitting the model-parameters to the data. Then the amplitude (an analytic function of the complex-valued energy) is analytically continued to the unphysical Riemann sheets where poles are allowed and are associated with resonances. For a pedagogical review, see Ref.~\cite{Mai:2025wjb}.

The vector, isovector $\rho(770)$-meson, has quantum numbers that can overlap with those of a $\pi\pi$ system. Thus, its parameters can be extracted from the scattering data of two pions. Indeed, phase-shift data for this process are available~\cite{Protopopescu:1973sh, Estabrooks:1974vu}. However, inconsistencies have been reported in the pole positions extracted from these data and the PDG values, obtained using data from a larger number of experiments, see e.g., Refs.~\cite{Ananthanarayan:2000ht, Colangelo:2001df, Garcia-Martin:2011iqs, Colangelo:2018mtw, Pelaez:2019eqa}. Thus, an open question remains whether a method different from the $\chi^2$-minimization can be used to mitigate possible inconsistencies in the data extracting the pole positions of the $\rho(770)$. We note that the $\rho(770)$-resonance has also been studied in lattice QCD~\cite{CS:2011vqf,Dudek:2012xn, Bali:2015gji,Wilson:2015dqa,Fu:2016itp, Alexandrou:2017mpi, Andersen:2018mau, ExtendedTwistedMass:2019omo, Guo:2018zss, Hu:2017wli, Hu:2016iij, Guo:2016zos, Fischer:2020yvw}. While beyond the scope of the current work, the connection of the SBI methodology to input from lattice QCD is not unthinkable, see, e.g., recent work~\cite{Salg:2025now} also building on alternative statistical tools.

The applications of this are not limited to the $\rho(770)$-resonance. Experimental data on $K^-p\to K^-p$ and other final states in the low-energy regime are suffering under large ambiguities~\cite{Sadasivan:2018jig,Pittler:2025upn} while also building a large bulk of an input required to access the universal parameters of the famous $\Lambda(1405)$ two-pole structure~\cite{Mai:2020ltx}. Similarly, ambiguities in the meson-electroproduction data can lead to incorrect extraction of baryon resonance parameters~\cite{Mai:2021vsw}. Furthermore, models for $\pi\pi$ scattering in the kinematic range covering the $\rho(770)$-resonance are also of high relevance for multi-hadron interaction. For example, understanding of the three-pion interaction requires  knowledge of the two-pion subsystem. For pertinent application ($a_1(1260)$ and $\omega(782)$) within the IVU approach~\cite{Mai:2017vot} see Refs.~\cite{Sadasivan:2020syi, Sadasivan:2021emk, Mai:2021nul, Feng:2024wyg, Yan:2024gwp} and for further related works Refs.~\cite{Doring:2025sgb, Garofalo:2022pux, Mai:2021lwb,Alexandru:2020xqf, Culver:2019vvu, Mai:2019fba, Mai:2019pqr,Culver:2019qtx, Mai:2018xwa, Mai:2018djl, Doring:2018xxx,Mai:2017wdv, Mai:2017bge,Mai:2017vot,Alexandru:2020xqf, Guo:2019ogp, Guo:2018ibd, JPAC:2019ufm, Albaladejo:2019huw, Mikhasenko:2019vhk}. 

This work is not the first to apply methods involving neural networks to study resonance phenomena. Methods similar to SBI are used in: Ref.~\cite{Ng:2021ibr} to predict the type of resonance; Ref.~\cite{Ng:2024xye} to  determine the pole position and pertinent Riemann sheets of the Pc(4312)+; Ref.~\cite{Santos:2024bqr} to predict the pole structure of P$\psi$N(4312)+; Ref.~\cite{Chavez:2025dro} to study the $\Lambda(1405)$ resonance; Ref.~\cite{Santos:2024bqr} to obtain the scattering length and effective range of the $X(3872)$ and the $T_{cc}+$; Ref.~\cite{Malekhosseini:2024eot} to predict meson mass and width based only on their quantum numbers; Refs.~\cite{Co:2024szz,Dersy:2023job} to analyze hidden charm pentaquarks. Refs.~\cite{Mapa:2025ihq, Co:2024bfl, Sombillo:2022sqf, Sombillo:2021ifs, Sombillo:2021rxv, Sombillo:2021yxe, Sombillo:2020ccg, Landay:2018wgf, Molina:2017anu, Petrellis:2022eqw, Bydzovsky:2021rog, Alghamdi:2023emm,Binosi:2022ydc} provide additional general methods for applying neural networks or other machine learning methods to the study of pole positions or analytic continuation. What distinguishes the present paper from previous work in applying deep learning to the study of resonances is an explicit focus on making predictions in the case of model misspecification.

This paper is structured as follows. Section~\ref{subsec:Methodology} describes the specific method of SBI that is applied in this work. Section~\ref{subsec:ToyExample} applies this method to a contrived example, for which the correct answer is known, in order to demonstrate the effectiveness of the method in comparison to $\chi^2$ minimization. The method is then applied to $\pi\pi$ scattering data. Section~\ref{subsec:SpecificPredictions} applies two separate models to two separate data sets, isolating effects of each model and data set. Section~\Ref{subsec:General} uses an SBI method to determine what model to use with the data and then to make a prediction for both data sets. Section~\ref{sec:Discussion} discusses reasons why the SBI method might outperform the method of $\chi^2$ minimization. The work is concluded in Sec.~\ref{sec:Conclusion}.

%%%%%%%%%%%%%%%%%%%%%%%%%%%%%%%%%%%%%%%%%%%%%%%%%%%%%%%
\section{Simulation Based Inference applied to model misspecification}
%%%%%%%%%%%%%%%%%%%%%%%%%%%%%%%%%%%%%%%%%%%%%%%%%%%%%%%

%%%%%%%%%%%%%%%%%%%%%%%%%%%%%%%%%%%%
\subsection{Methodology}
\label{subsec:Methodology}
%%%%%%%%%%%%%%%%%%%%%%%%%%%%%%%%%%%%

This paper presents an alternative method for fitting a function, which can be written as $f(x; \vec a)$ with $x$ and $\vec{a}$ being an independent variable and model parameter list, respectively. The problem of fitting involves comparing this function to data to determine the most probable free parameters. This is often done by minimizing a loss function such as the $\chi^2$ statistic, which (if correlations in data are neglected) is given by
%%%%%%%%%%
\begin{align}
    \chi^2=\sum_i^{N_{data}} \frac{(f(x_i; \vec a)-y_i)^2}{\sigma_i^2}\,,
\end{align}
%%%%%%%%%%
where $x_i$ are independent measurements, and $y_i$ denote  the dependent measurement with uncertainties $\sigma_i$. This well-known formula can be used to find the most probable parameters $\vec{a}$ given the measured data $y_i$ under the assumption that the data are normally distributed about the function for some values of $\vec{a}$, and the prior uniform probability distribution of fit parameters.  

%%%%%%%%%%%%%%
%%%%%%%%%%%%%%
\begin{figure*}[t]
    \includegraphics[width=0.8\linewidth]{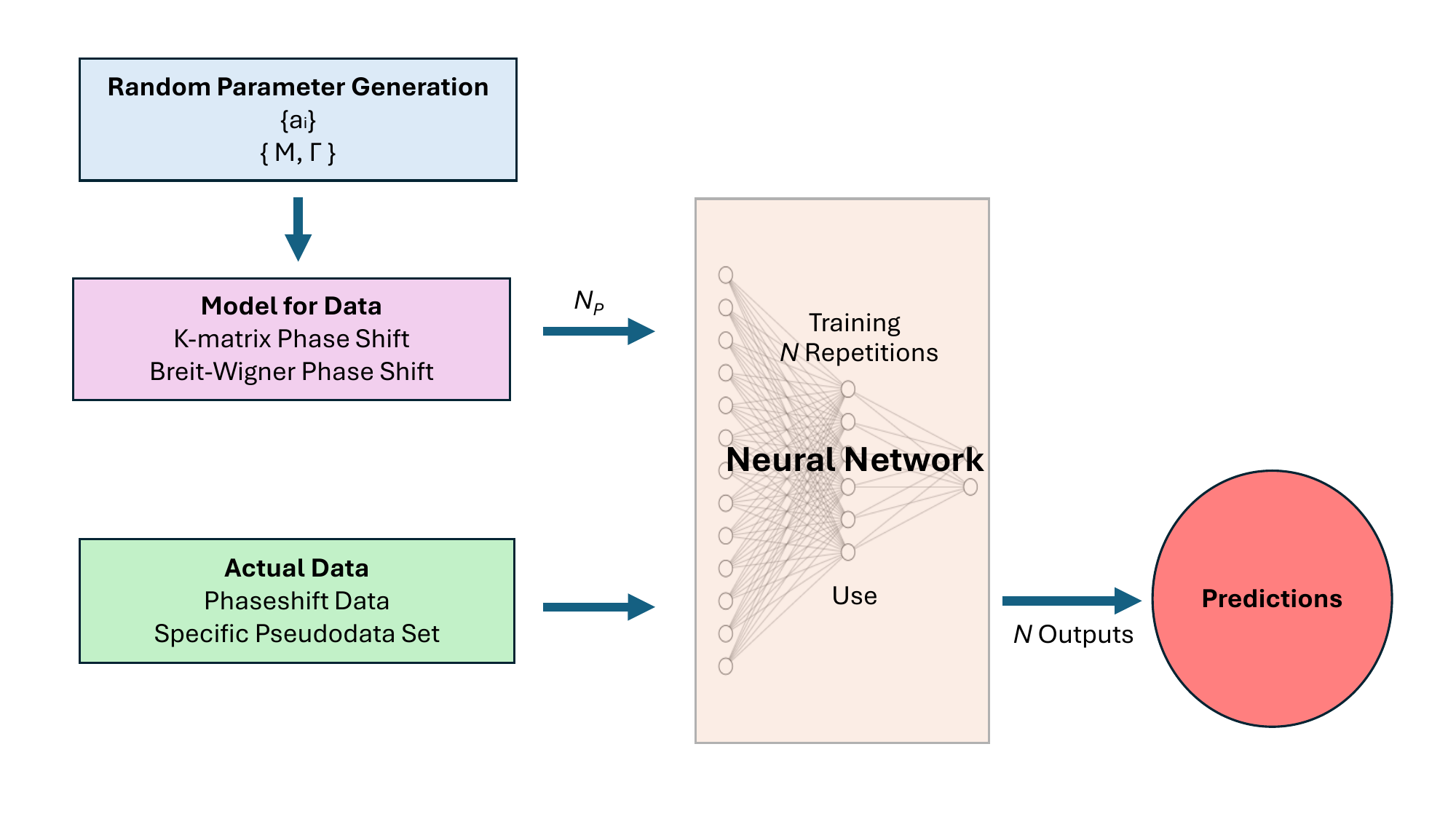}
    \caption{A diagrammatic depiction of the data pipeline in the SBI method employed in this work. In each of the boxes on the left the middle/bottom row refer to the studied case of actual $\pi\pi$ scattering data and the toy example, respectively. 
    }
\label{fig:DataPipeline}
\end{figure*}
%%%%%%%%%%%%%%
%%%%%%%%%%%%%%

In the case of model misspecification, meaning that no values of $\vec{a}$ exist for which the data are normally distributed about the function, the most accurate parameters (given the data) may not be the parameters corresponding to the minimum $\chi^2$. SBI relies on different assumptions and therefore, in certain situations, could give more accurate probability distributions in these situations. Because there are multiple ways that SBI can be implemented, we describe the method employed in this work below, also visualizing the data pipeline in Fig.~\ref{fig:DataPipeline}.
%%%%%%%%%%%%%%
%%%%%%%%%%%%%%
\begin{enumerate}[label=\arabic*., leftmargin=0pt, labelsep=1em, itemindent=!, align=left, listparindent=0pt, labelwidth=*, itemsep=1ex]
    %%%
    \item Sets of parameters $\vec{a}_j$ are randomly generated. The range and probability distribution for this must be carefully chosen to minimize introduced bias. If most of the generated $\vec{a}_j$ lead to a certain feature, the SBI method might predict this feature, whether or not the data contain evidence for this feature. The distribution used to generate these parameters can be thought of as the Bayesian prior distribution for the parameters.
    %%%
    \item For each of these parameter sets, the function $f$ is used to generate a set of pseudodata, $x^p$, $y^p$ and $\sigma^p$, where $x^p_i=x_i$ and $\sigma_i^p=\sigma_i$ for provided independent variable and corresponding uncertainties. In other words, the independent variable and the uncertainty are the same for the pseudodata as for the actual data.  The dependent variable is determined as $y_i^p=R(f(x_i),\sigma_i)$, i.e., a normally distributed random variable with a mean $f(x_i)$ and standard deviation $\sigma_i$. When done this way, the method works for a fixed set of data. If new data is added, new values of $x_i$ must be used. In principle, this method can also be extended to assess the uncertainty on the independent variable, $x_i$. This can be of large relevance, e.g., for negative-strangeness meson-baryon systems~\cite{Sadasivan:2022srs, Sadasivan:2018jig, Guo:2023wes, Bruns:2022sio, Mai:2018rjx, Cieply:2016ull, Cieply:2016jby, Mai:2014xna, Feijoo:2018den}, where experimental data not only has large error bars on the measured cross sections but also on the energy variable. In these circumstances, pseudodata for the independent variable can be generated with $x_i^p=R(x_i,\sigma_{xi})$, where $\sigma_{xi}$ is the uncertainty on $x_i$.
    %%%
    \item If necessary, only data points with the feature being studied are retained. For instance, in our example of $\rho(770)$-resonance, we only retain data points with physical poles in a realistic range. Clearly, this step can introduce bias similar to the bias described in Step 1. For instance, if a neural network that is only trained on data corresponding to reasonable $\rho$ poles returns a $\rho$ pole that is no more accurate than the average of the training data, this is not evidence that the neural network has learned the data set. This will be discussed in detail in \cref{appc:DataGeneration}. The number of the remaining pseudodata sets is referred to in the following as $N_P$.
    %%%
    \item The pseudodata is used to train a neural network. The values $y_i^p$ are used as the inputs of the neural network and the randomly generated $\vec{a}$ used as outputs. A variety of loss functions could be used for this training, however,  the most common is the mean squared error, which is also employed in this work.
    %%%
    \item Once the neural network is trained, $y_i$ from the actual data are given as inputs and the outputs, predictions for $\vec a$, are recorded.
    %%%
    \item Steps 4. and 5. are repeated $N$ times. This is necessary because there is some randomness in the selection of batches of $\{y^p,\vec a\}$ while training the neural network. The output predictions for  $\vec{a}$ are stored.
    %%%
    \item The average values of the model parameter list obtained in step 6 give the best fit predictions $\bar a$. 

    Assuming a symmetric normal distribution of the values, the uncertainty of the average of the neural networks is obtained through standard deviation as $\bar{a}_i\pm \sigma(a_i)/\sqrt{N}$ for each model parameter $a_i$. This term accounts only for the uncertainty of the average of the neural network in the limit of a large number of trials, not for 
    the bias that might occur if this average does not converge to the true value. 
     To our knowledge, there is no general method that accurately estimates the value of this bias in all cases. This can present a real challenge for the use of SBI. Due to this concern, confidence regions for the SBI predictions should be treated with less certainty than equivalent confidence regions obtained through a $\chi^2$ minimization.
    However, in this paper, we have focused on the case of model misspecification, for which, $\chi^2$ minimization also returns biased results with no clear way to estimate the magnitude of the bias. Thus, this challenge is not unique to SBI.

\end{enumerate}
%%%%%%%%%%%%%%
%%%%%%%%%%%%%%

%%%%%%%%%%%%%%%%%%%%%%%%%%%%%%%%%%%%
\subsection{Toy Examples}
\label{subsec:ToyExample}
%%%%%%%%%%%%%%%%%%%%%%%%%%%%%%%%%%%%

We begin with two toy examples to demonstrate the SBI. In the first, model misspecification is created by intentionally modifying several data points to prevent a reasonable fit, and in the second, model misspecification is created by generating synthetic data with non-Gaussian uncertainty but not including information about this probability distribution in the $\chi^2$ minimization or SBI fit. In both examples we use a Breit-Wigner like parametrization of the scattering amplitude
%%%%%%%%%%%%%%
\begin{align}
    \label{eq:Toy}
    &T(E)=\frac{\sqrt{k}}{(E^2-M^2)-iM\Gamma}\,
    %&\delta(E)=\tan^{-1}\left(\frac{\Im\,T(E)}{\Re\,T(E)}\right) \\
\end{align}
%%%%%%%%%%%%%%
with a resonance mass $M$, width $\Gamma$ and energy $E$. Using this parametrization, we generate unmodified synthetic data, i.e., phase-shift $\delta(E)=\arctan(\Im\,T(E)/\Re\,T(E))\,$ that would be plausible for a specific value of $M$ and $\Gamma$.

%%%%%%%%%%%%%%%%%%%%%%%%
\subsubsection{Example 1 (outliers)}
%%%%%%%%%%%%%%%%%%%%%%%%

\begin{figure*}[t]
    
\includegraphics[width=0.45\textwidth]{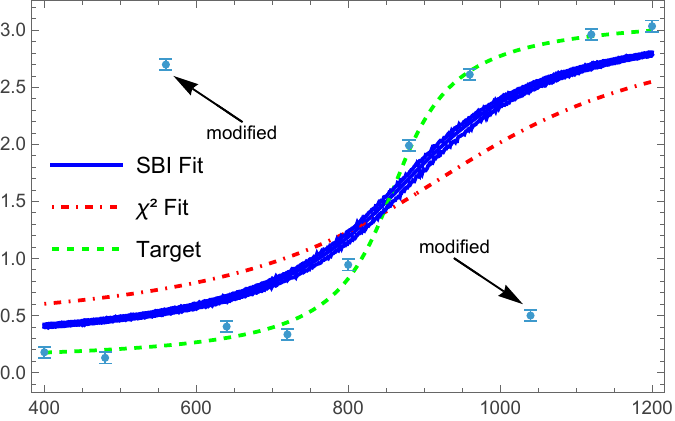}
\hspace{0.05\textwidth}
\includegraphics[width=0.45\textwidth]{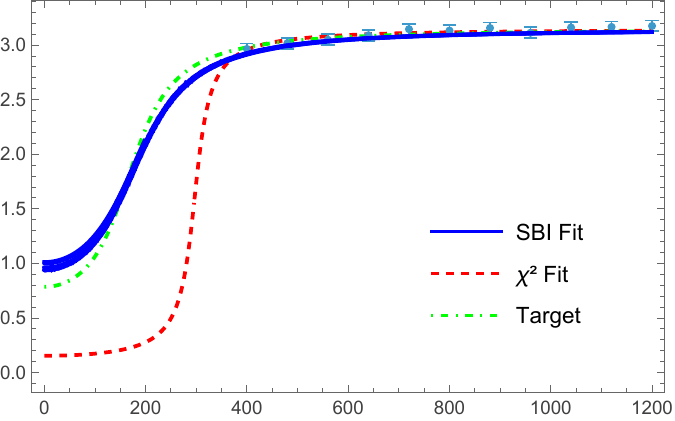}

 \caption{Left: Phase-shift data with two points that are intentionally modified to create a case of extreme model misspecification in comparison to the function that gives the minimum $\chi^2$  and the SBI result including confidence region. The SBI curve lies much closer to the unmodified data than the curve computed through $\chi^2$ minimization.
 Right: A similar comparison of the SBI method to the neural network method. In this case the synthetic data is not modified but instead is generated with a non-Gaussian distribution. The SBI neural network is trained on only Gaussian-generated pseudo data so that it does not have an advantage.}
\label{fig:ToyPlots}
\end{figure*}

The first toy example is intentionally chosen to provide an extreme case to visualize the SBI method when dealing with outliers. This example is performed selecting the parameters  $M=857\MeV$ and $\Gamma=119\MeV$ and then intentionally changing the data (modified data) such that no values of $M$ and $\Gamma$ could plausibly have generated the data. Specifically, 11 unmodified points are generated for values of $E$ ranging from 400~MeV to 1200~MeV in increments of 80~MeV, adding a normally-distributed random variable with a standard deviation of 0.05 to $\delta(E)$.  After the unmodified data is calculated, the points are modified such that $\delta(E=560\MeV)=2.7$ and $\delta(E=1040\MeV)=0.5$ as depicted in Fig.~\ref{fig:ToyPlots}. Note that this stark example is chosen as a first test contrasting $\chi^2$ vs SBI methodology, only. While another case is discussed below, we note that dealing with outliers is, indeed, an important aspect in many modern hadron spectroscopy. Partially, this is due to outdated, old data sets or simple typos in data basis, see, e.g., the discussion in Refs.~\cite{Mai:2021vsw,Mai:2021aui,Mai:2023cbp,Doring:2025sgb}. Methods mitigating this issue range from manual checks to data pruning~\cite{NavarroPerez:2014ihw, NavarroPerez:2013mvd, Landay:2018wgf}.

%%%%%%%%%%%%%%
%%%%%%%%%%%%%%
%%%%%%%%%%%%%%
%%%%%%%%%%%%%%

The probability that the modified data could have been generated from any values of $M$ and $\Gamma$ is vanishingly small. The usual $\chi^2$ minimization can, still, be applied to the modified data. The best fit returns values of $M$ and $\Gamma$ provided in the upper half of Tab.~\ref{tab:ToyParameters} while the corresponding phase-shift is shown in the left side of Fig.~\ref{fig:ToyPlots}.

%%%%%%%%%%%%%%
%%%%%%%%%%%%%%
\begin{table}[t]
    \caption{A comparison of the free parameters obtained through $\chi^2$ minimization and SBI.  The upper half shows the values using modified data and the lower half shows the values for the non-Gaussian synthetic data. The parameters used to generate the unmodified data are given in the second column. 
    %While the $\chi^2$ mass is slightly closer to the unmodified mass than the SBI mass is, the SBI width is more than an order of magnitude closer to the unmodified width. If the goal is to obtain the parameters closest to the unmodified data, the SBI method outperforms the $\chi^2$ method. 
    \label{tab:ToyParameters}
    }
    \begin{tabular}{|p{3.2cm}|l|l|l| } 
    \hline
    Modified &Generating&$\chi^2$&SBI\\ 
     \hline
     $M~[{\rm MeV}]$&857.0&$889.2 \pm 6.5$& 861.3 $\pm$ 5.0 \\ 
     $\Gamma~[{\rm MeV}]$&$119.0$&$482.6\pm 17.2$&293.5 $\pm$ 6.5 \\ 
     \hline
     \hline
     Non-Gaussian&Generating&$\chi^2$&SBI\\
     \hline
     $M~[{\rm MeV}]$&150.0&$291.6 \pm 62.0$&  148.7 $\pm$ 1.2\\ 
     $\Gamma~[{\rm MeV}]$&150.0&$\phantom{0}56.7 \pm 52.3$& 207.7 $\pm$ 2.8 \\ 
     \hline
    \end{tabular}
\end{table}
%%%%%%%%%%%%%%

Simulation Based Inference is also employed, with its predictions shown the right column of Tab.~\ref{tab:ToyParameters}. There, we randomly generate $N_P=10^6$ points of training data with values of $M/\MeV$ uniformly distributed between 400 and 1200 and values of $\Gamma/\MeV$ uniformly distributed between 0 and 500. The pseudodata is passed to a neural network.  Specifically, they include 4 hidden layers, with 40, 30, 20, and 10 nodes, respectively.  Each node in each layer has a \texttt{relu} activation function. The output layer has two nodes with the leaky \texttt{relu} activation function. The mean squared error loss function is used. The neural networks are run for 1000 epochs or until they have been run for 20 epochs without improving the validation loss. In almost all cases, the latter occurs, indicating that further epochs would not improve the validation loss. The batch size chosen for the training process is 2000. The neural network is trained with the Adam optimizer. We note that the \texttt{relu} activation function, the Adam optimizer, and the mean squared error loss function are standard and have not been fine-tuned for a specific problem at hand. The leaky \texttt{relu} for the output layer was chosen because more common activation functions cannot return negative values.  The number of nodes and layers was chosen somewhat arbitrarily. To assess this possible source of bias, we have performed numerical tests described in the App.~\ref{App:Architecture}, showing that the size and architecture of the neural network is reasonable. We employ the same neural network architecture across all cases.

Taking a closer look on the results in Tab.~\ref{tab:ToyParameters}, we note that the SBI and $\chi^2$ method give an $M$ that is similarly close to the original $M$ used to generate the unmodified data. At the same time, the value of $\Gamma$ in the SBI method is closer by more than an order of magnitude than the value from the $\chi^2$ method to the true value. The confidence region of $\delta(E)$ predicted by SBI shown in the blue shaded area of Fig.~\ref{fig:ToyPlots}. While this region does not cover all of the data points, it is much closer than the $\delta(E)$ calculated with the parameters obtained through the $\chi^2$ minimization. This toy model demonstrates that, under model misspecification the SBI can outperform $\chi^2$ minimization providing more accurate predictions of the underlying parameters.

Note that, the neural network with a mean squared error loss function is not minimizing the same quantity as the $\chi^2$. Even though both quantities are a sum of squares, the $\chi^2$ statistic is the sum of the squared difference between the predicted and measured values of the phaseshift, $\delta$, whereas the mean squared error is the squared difference between the predicted and actual values of the fit parameters, $\vec a$. Thus, the SBI prediction for the phase-shift should not be expected to converge with the $\chi^2$ prediction. 

In this toy example, the exact nature of the model misspecification is known. Thus, a reasonable alternative strategy could include, removing the visibly inaccurate data points as outliers or using different loss function, rather than the $\chi^2$ function to reduce reliance on a few points. However, the strength of using SBI is that it did not have to identify why the data or model was flawed. In many cases of real application, the flaws that cause model misspecification are not identifiable. SBI might be used in some of these situations to obtain more accurate results even without identifying them.

%%%%%%%%%%%%%%%%%%%%%%%%
\subsubsection{Example 2 (non-Gaussian errors)}
%%%%%%%%%%%%%%%%%%%%%%%%

As a second toy example we consider a case where the data used to obtain final results from the neural network are not modified but are generated from a non-Gaussian probability distribution. Specifically, random noise is added to the phase-shift generated from an exponential distribution given by $e^{-(x-1)\lambda}\lambda$ for $x>0$, with $\lambda=0.05$. This distribution has mean of 0 and a standard deviation of 0.05 just as in the previous toy model. However, due to the asymmetry of this distribution, more data points lie above the curve than below however, the points that do lie below are more likely to be far below, which gives them a disproportionate influence on the $\chi^2$. Thus, this is a situation of model misspecification even though the data have not been modified.

In this case, we choose the "true" parameters used to generate the data to be $M=150$ and $\Gamma=150$. The data are generated for the same values of $E$ as the previous case. We have chosen different generating parameters for this toy model and the previous one because after trying several non-Gaussian distributions for the generating pole parameters of the previous toy-model (chosen to provide a good visualization of the SBI method) we were unable to create a case of model misspecification. This does not, however, mean that in general such distribution does not exists. We note that this situation still has important physical relevance, e. g., for subthreshold resonances, such as $\Lambda(1405)$~\cite{Mai:2020ltx} or $T_{cc}^+(3875)$ mentioned above.

The pseudodata used to train the neural network is still Gaussian, because otherwise the knowledge of target distribution is needed. Of course, the SBI method would obtain more accurate results if it were trained on data generated with the same probability distribution as the one used for the data in the prediction, but we choose to train the neural network on the Gaussian data to make the comparison more even. The $\chi^2$ statistic  is constructed without knowledge of the probability distribution of the data generated. We work with an SBI method that has no additional knowledge. The parameters for this method are given in Tab.~\ref{tab:ToyParameters} and the curves are shown in Fig.~\ref{fig:ToyPlots}. As in the case of the previous toy example, the SBI method predicts parameters that are closer to the ones used to generate the data than the ones predicted by the $\chi^2$ method.

%%%%%%%%%%%%%%
%%%%%%%%%%%%%%
\begin{table*}[t]
    \caption{The pole positions of the  $\rho(770)$ from SBI and $\chi^2$ methods for data sets from Ref.~\cite{Protopopescu:1973sh} (Protopopescu) and Ref.~\cite{Estabrooks:1974vu} (Estabrooks). Two and three subtractions ($n$) are used in both methods. Several reference values from the PDG~\cite{ParticleDataGroup:2024cfk} are provided in the lower table for comparison. All pole positions are visualized in Fig.~\ref{fig:TemperatureRatios}.}
    \label{tab:chi2}
    \renewcommand{\arraystretch}{1.4} % Default is 1.0
    \begin{tabular}{|l|l||
        l|l|l|l||
        l|l|l|r|
        l|}
        \multicolumn{2}{c}{}&\multicolumn{4}{c}{SBI}&\multicolumn{5}{c}{$\chi^2 $}\\
        \hline
        Data & $n$ 
        & $a_0\cdot10^3$ & $a_1\cdot10^3$ & $a_2\cdot10^3$ 
        & $E^*[{\rm MeV}]$
        & $a_0\cdot10^3$ & $a_1\cdot10^3$ & $a_2\cdot10^3$ & $\chi^2_{\rm dof}$
        & $E^{*}[{\rm MeV}]$ \\
        \hline
        \hline
        Estabrooks & 2 
        & $-444$ & $+14.2$ &  & $766.3(1.8)-i\,79.0(0.7)~~$
        & $-460$ & $+15.6$ &  & $12.8$ 
        & $750.3(0.6)-i\,69.8(0.5)$\\
        \hline
        Protopopescu & 2 
        & $-454$ & $+14.9$ &  & $763.7(1.0)-i\,75.5(0.3)$
        & $-467$ & $+15.7$ &  & $3.7$
        & $755.6(0.8)-i\,70.4(0.6)$ \\
        \hline
        Estabrooks & 3 
        & $-188$ & $-3.22$ & $+0.41$ & $769.6(2.9)-i\,71.8(0.3)$
        & $-223$ & $+0.93$ & $+0.39$ & $3.6$
        & $766.8(1.5)-i\,74.0(0.7)$ \\
        \hline
        Protopopescu & 3 
        & $-368$ & $+8.90$ & $+0.20$ & $772.3(2.1)-i\,78.7(0.6)$
        & $-319$ & $+6.30$ & $+0.25$ & $0.4$
        & $762.8(1.2)-i\,78.5(1.1)$ \\
        \hline
        All & 3 
        & $-187$ & $-3.17$ & $+0.42$ & $772.4(2.3)-i\,75.4(0.4)$
        & $-290$ & $+4.10$ & $+0.29$ & $3.1$
        & $763.7(0.9)-i\,\,76.9(0.6)$ \\
        \hline
    \end{tabular}
    %%%%%%%%%%%%%
    ~\\~\\
    \begin{tabular}{|l|l|}
        \multicolumn{2}{c}{Reference/PDG values}\\
        \hline
        &$E^*=M-i\,\Gamma/2\,[{\rm MeV}]$ \\
        \hline
        $e^+ e^-$ &$775.3(0.3)-i\,73.9(0.5)$ \\
        \hline
        Photoproduction &$769.0(1.0)-i\,75.9(1.3)$  \\
        \hline
        Other Reactions &$769.0(0.9)-i\,75.5(0.9)$\\
        \hline
    \end{tabular}
\end{table*}
%%%%%%%%%%%%%%
%%%%%%%%%%%%%%

%%%%%%%%%%%%%%%%%%%%%%%%%%%%%%%%%%%%%%%%%%%%%%%%%%%%%%%%
\section{Application to $\pi\pi$ scattering and the $\rho$(770)}
%%%%%%%%%%%%%%%%%%%%%%%%%%%%%%%%%%%%%%%%%%%%%%%%%%%%%%%%

As discussed before, the properties of the $\rho (770)$-resonance can be accessed through  $\pi\pi$-scattering data. As a representative example, we employ in this paper fits to the $\pi\pi$-scattering data using a K-matrix like approach reading
\begin{align}
        % \delta(E)&=\tan^{-1}\left(\frac{\text{Im}\,T(E)}{\text{Re}\,T(E)}\right) \nonumber, \\
        &T(E)=\tilde v(k_\text{cm})\frac{1}{K_n^{-1}(E)-\Sigma(E)}\tilde v(k_\text{cm})\,,\nonumber\\
        &K_n^{-1}(E)=M_\pi^2\sum_{i=0}^{n-1} a_i \left(\frac{E^2}{M_\pi^2}\right)^i\,,   \nonumber\\
        &\Sigma(E)=\int\limits_0^\infty\frac{ dk\,k^2}{(2\pi)^3}\,\frac{1}{2E_k}\left(\frac{E^2}{4E_k^2}\right)^n\frac{\tilde v^2(k)}{\sigma-4E_k^2+i\epsilon}\,,
\label{eq:tau-infinite}
\end{align}
used, e.g., in Ref.~\cite{Sadasivan:2021emk}. Here, $k_{\rm cm}=\sqrt{E^2/4+M_\pi^2}$, $\tilde v(k)=\sqrt{16\pi/3}k$, $E_k=\sqrt{k^2+M_\pi^2}$, while $M_\pi$ denotes the pion mass. These models can use any number of subtractions ($n$) which defines the number of free parameters of the model ($a_i$). When the free parameters of these models (e.g., $n=2$ or $n=3$ subtractions) are fit to the experimental data from Refs.~\cite{Protopopescu:1973sh} and~\cite{Estabrooks:1974vu} using $\chi^2$ minimization the predicted pole positions differ significantly from the PDG values~\cite{ParticleDataGroup:2024cfk} including larger data sets. This problem is not unique to the K-matrix but was also discussed in a broader context~\cite{Ananthanarayan:2000ht, Colangelo:2001df, Garcia-Martin:2011iqs, Colangelo:2018mtw, Pelaez:2019eqa} indicating the presence of certain inconsistencies in these data. This section covers the application of SBI to $\pi\pi$ scattering to make predictions for the pole position of the $\rho(770)$.  This is done for each data set and model separately in Sec.~\ref{subsec:SpecificPredictions} and all together in Sec.~\ref{subsec:General}.

%%%%%%%%%%%%%%%%%%%%%%%%%%%%
\subsection{Predictions for Individual Data Sets and Models}
\label{subsec:SpecificPredictions}
%%%%%%%%%%%%%%%%%%%%%%%%%%%%

%%%%%%%%%%%%%%
%%%%%%%%%%%%%%
\begin{figure*}[t]
\includegraphics[width=0.7\textwidth,trim=90 50 90 80]{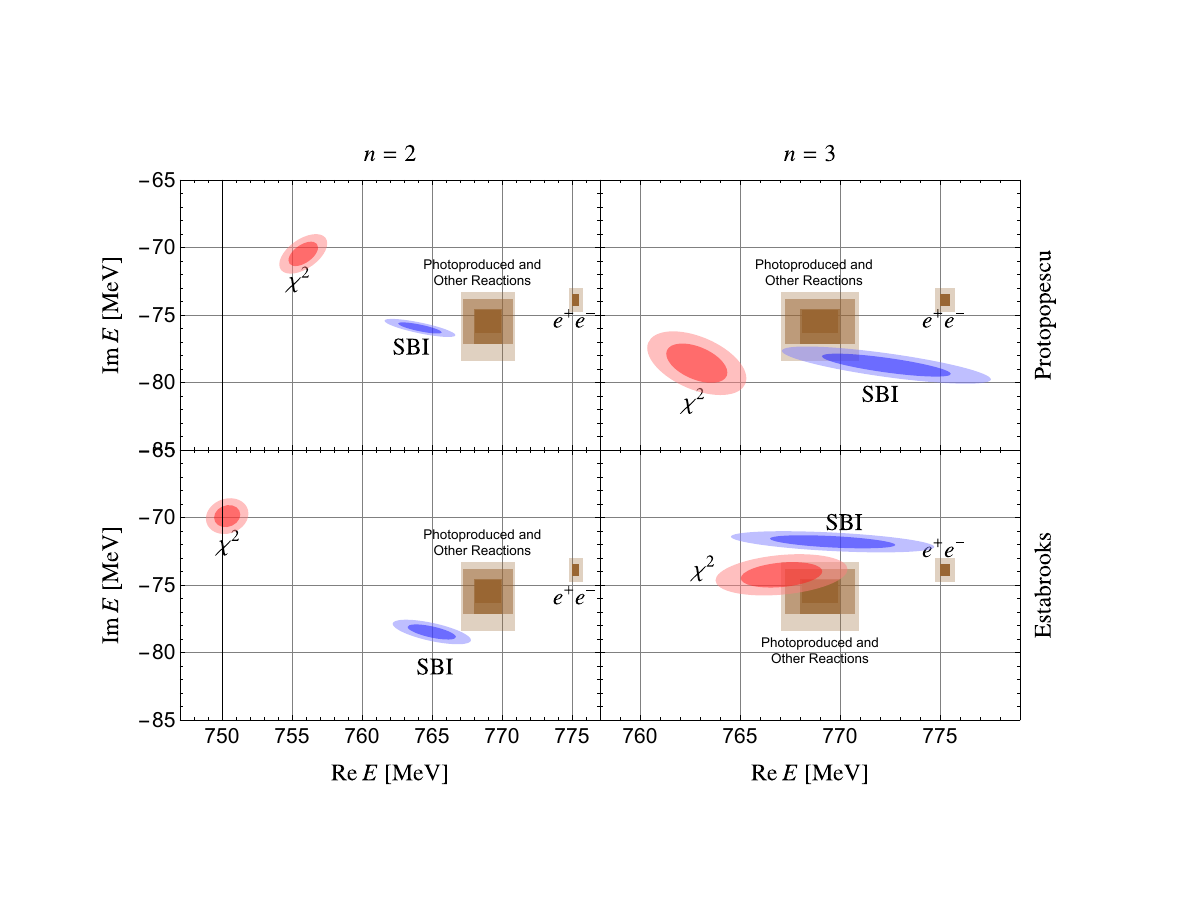}
\caption{
The pole position confidence regions of the $\rho(770)$ for data sets from Ref.~\cite{Protopopescu:1973sh} (Protopopescu) and \cite{Estabrooks:1974vu} (Estabrooks), using two and three subtractions. The red rectangles (1 and 2 standard deviations) give the confidence region obtained through $\chi^2$ minimization and the blue that of the SBI. The brown rectangles give the PDG values for various types of experiments which serve as a benchmark. In these cases, the SBI method makes predictions for the $\rho$-pole that as close or closer to the reference values than the $\chi^2$ method. 
It should, however, be noted that each case is related to two others so this result is not as strong as if all four cases were completely independent. Furthermore, the confidence regions given for the SBI and $\chi^2$ do not account for bias in the prediction and thus should not be treated with the same credence as predictions obtained in the absence of model mispecification. }
\label{fig:TemperatureRatios}
\end{figure*}
%%%%%%%%%%%%%%
%%%%%%%%%%%%%%

In this section, we apply two and three subtraction models (c.f. $n$ in \cref{eq:tau-infinite}) to data from Refs.~\cite{Protopopescu:1973sh} or ~\cite{Estabrooks:1974vu} to which we refer to as Protopopescu data and Estabrooks data after the respective first author of each publication. This defines four different cases, allowing us to assess the  effects that might be specific to one model/data set.

The best fit parameters with $\chi^2$ minimization are determined and shown including the extracted pole positions in the right part of the \cref{tab:chi2}. Left part of the same table collects the results of the corresponding SBI method.
For the $\chi^2$ method, the uncertainties are calculated with bootstrap resampling, while the pseudodata are generated following the method described in~\cref{subsec:Methodology} with some details relegated to \cref{appc:DataGeneration}.
The bottom of \cref{tab:chi2} gives the reference values taken from the PDG tables~\cite{ParticleDataGroup:2024cfk} where larger range of data sets and factors were used. We emphasize that our goal here is not to improve on theoretical tools (see, e.g., Refs.~\cite{Hoferichter:2023mgy, Garcia-Martin:2011nna, Pelaez:2004xp, Colangelo:2001df}) but rather to study possible alternatives to the usually employed $\chi^2$ methodology. The positions for each case for both the $\chi^2$ method and the SBI method, as well as the PDG pole positions are visualized in \cref{fig:TemperatureRatios}.

In all four cases, the SBI pole position is closer to the pole position from the most precisely determined $e^+e^-$ data. 
%In three of the four cases, the SBI is closer to all predictions of the data than the PDG. In the Estabrooks $n=3$ case, the $\chi^2$ central predicted value is slightly closer to the photoproduction data and data from other reactions. %However, the 95\% confidence region for the SBI prediction in this case overlaps with the equivalent confidence regions for all three PDG reference values. This indicates that being closer to the PDG values for the photoproduction data and other data might not even be a sign of greater accuracy. 

Overall, it is reasonable to claim that in all four cases, the SBI method leads to pole predictions that are as close or closer to the reference values than the method of $\chi^2$ minimization.  Note, however, that because the different models are partly related and because there are only two models and two data sets, this result should not be considered to be as strong as if four completely independent cases were employed.

%%%%%%%%%%%%%%
%%%%%%%%%%%%%%
\begin{figure*}[t]
    \includegraphics[width=0.41\linewidth]{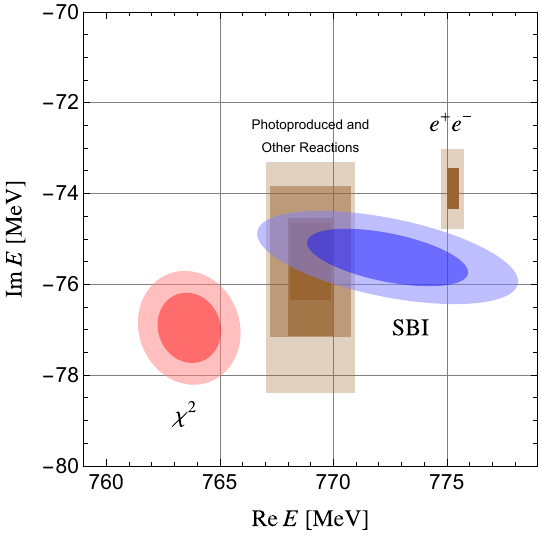}
    ~~~~~~~~~~
    \includegraphics[width=0.41\linewidth]{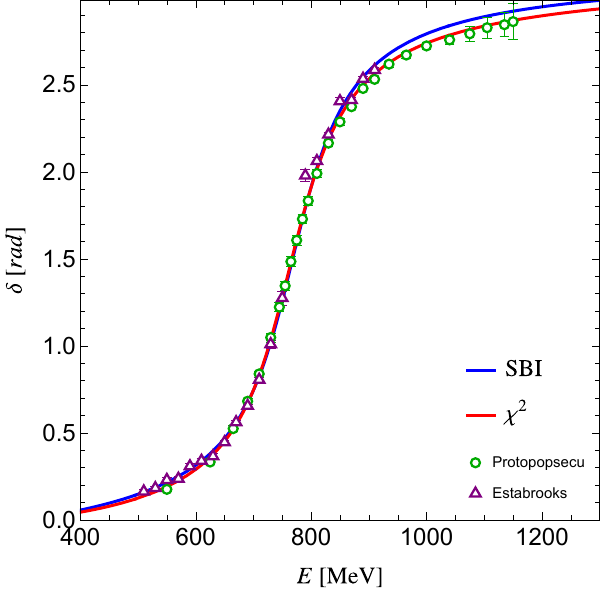}
    \caption{Left: A comparison of the SBI method (blue) and the $\chi^2$ method (red) for the pole positions. The SBI pole position is closer to all PDG data sets than the $\chi^2$ minimization.
    Right: A comparison of the phase-shifts for the SBI method  and the $\chi^2$ method  with the  Protopopsecu~\cite{Protopopescu:1973sh} and  Estabrooks~\cite{Estabrooks:1974vu} data. The best fit $\chi^2_{\rm dof}$ value and further details can be found in \cref{tab:chi2}.}
    \label{fig:AllDataPlot-PhaseShiftDataPlot}
\end{figure*}
%%%%%%%%%%%%%%
%%%%%%%%%%%%%%

%%%%%%%%%%%%%%%%%%%%%%%%%%%%
\subsection{Combined Fit and Classifier Network}
\label{subsec:General}
%%%%%%%%%%%%%%%%%%%%%%%%%%%%

In this section, we discuss how to incorporate information for all four cases. The two data sets can be combined into a single data set, but both different models cannot be combined -- one or another must be chosen. The method we use to determine the best model to use to fit the data is similar to the SBI method used in this work, see \cref{subsec:Methodology}. A large number of pseudodata sets are generated from models with various number of subtractions (c.f., $n$ in \cref{eq:tau-infinite}). Only pseudodata points that have realistic pole positions are retained as discussed in \cref{appc:DataGeneration}. A neural network is then trained to predict the number of subtractions that generated the pseudodata. Finally, the experimental data is given to the neural network returning an estimate of the probability that this data was generated from a given model assuming it was generated from such model types. This classifier neural network uses the softmax activation function for the final node. The network is trained using a cross-entropy loss function. The estimated probabilities $P(n)$ with uncertainties in the parenthesis read 
%%%%%%%%%%%%%%
\begin{align}
    P(1)&=0.00067(00003)\,,   \nonumber \\
    P(2)&=0.06482(00095)\,,   \nonumber \\
    P(3)&=0.86657(00122)\,,   \label{tab:probabilities} \\
    P(4)&=0.06791(01004)\,,   \nonumber\\
    P(5)&=0.00001(00000)\,.    \nonumber
\end{align}
%%%%%%%%%%%%%%
These results indicate that, if the actual data were generated from a K-matrix model, three subtractions is most likely to have generated it, which we use in the following. Note that these predictions are not for a specific set of $a_i$ but for any values of $a_i$.

The SBI predictions using both data sets are given in bottom row of \cref{tab:chi2} together with the $\chi^2$ parameters, while the predicted pole positions are also depicted in the left panel of \cref{fig:AllDataPlot-PhaseShiftDataPlot}. Right panel of \cref{fig:AllDataPlot-PhaseShiftDataPlot} shows the model phase-shift predictions in comparison to the data. 
The degree of model misspecification in this case is much lower than in the Toy Example which was specifically chosen as an extreme case. It may not be "visible by eye," however, the $\chi^2_{\rm dof}$ for this case is 3.1 (given in ~\cref{tab:chi2}). For this situation, the $\chi^2$ corresponds to a $p$-value of $0.17\times 10^{-3}$ indicating that model misspecification is clearly present. The red $\chi^2$ curve lies closer to the data than the SBI method. This fact is not surprising because the $\chi^2$ statistic is a measure of how far the curve is from the data and, therefore, the curve corresponding to the minimum $\chi^2$ will be the one closest to the data. Nevertheless, the SBI curve makes a prediction for the $\rho$ pole position that is in better agreement with PDG data sets (as shown in the left of~\cref{fig:AllDataPlot-PhaseShiftDataPlot}). This situation can be compared to the toy example in which the $\chi^2$ curve was closer (as measured by squared distance) to the modified data, but the SBI method made a prediction for the pole position that is in closer agreement with the model used to generate the majority of the data points. This indicates that possible ambiguities in the data are systematically better handled in the SBI method.

%%%%%%%%%%%%%%%%%%%%%%%%%%%%%%%%%%%%%%%%%%
\subsection{Training Data Generation and Tests for Bias}
\label{appc:DataGeneration}
%%%%%%%%%%%%%%%%%%%%%%%%%%%%%%%%%%%%%%%%%%

We generate pseudodata as described in Step 1 in \cref{subsec:Methodology}. The parameters $a_0$, $a_1$, $a_2$ are generated randomly with a uniform distribution within the following ranges $[-\GeV^2/M_\pi^2,\GeV^2/M_\pi^2]$, $[-1,1]$, $[-M_\pi^2/\GeV^2,M_\pi^2/\GeV^2]$, respectively. These ranges are well beyond the values that would generate reasonable $\rho(770)$-resonance poles. 

Next, as described in Step 3 of Sec.~\ref{subsec:Methodology}, we include only parameters which lead to a pole position within the largest possible square centered on the pole-position (obtained from the $\chi^2$ method) that is below the real axis. In other words, if the $\chi^2$ determined pole from the actual data has a position $E^*=a-ib$, we retain pseudodata points if the corresponding $\chi^2$ determined pole $E^*_P=x-iy$ fulfills $a-b<x<a+b$ and $2b<y<0$. Such a method centers the range of allowed pole positions on the $E^*$ but does not require addition choices of ranges. This pseudodata is generated until $N_P=10^6$ points of training data are obtained. This step is necessary because most randomly generated parameters do not lead to any physical pole and only approximately 2\% lead to a pole in the range in which we retain data. If this step were not performed, the neural network with the lowest loss function would be one that made the best predictions on the majority of the training data, which did not have poles. We emphasize again that the goal of this work is not to provide a universal neural network trained to assess any scenario but rather to provide an alternative method (SBI) to the usual $\chi^2$ minimization for extracting pole positions from data which includes ambiguities. In principle, anchoring the training data to the $\chi^2$ can effect the final predictions. However, this step ensures that, if the SBI method makes a more accurate prediction than the $\chi^2$ method, it is not because it is trained on data that is closer to known position.

%%%%%%%%%%%%%%
%%%%%%%%%%%%%%
\begin{table*}[bth]
    \caption{The average of the pole position calculated from each point in the pseudodata used to train the neural network. The right three columns give the distance in MeV between the pole positions in this table and the pole positions predicted by the SBI method and $\chi^2$ minimization method in Tab.~\ref{tab:chi2}. }
    \label{tab:PseudoComparison}
    \renewcommand{\arraystretch}{1.4} % Default is 1.0
    \begin{tabular}{|l|l||l||l|l|l|l|}
        \hline
        Data&$n$ & pseudodata $E^*\,[{\rm MeV}]$~~~~~~~~~~~&
        Distance $\chi^2$ to pseudodata &Distance pseudodata to SBI&Distance SBI to $\chi^2$ \\
        \hline
        \hline
        Estabrooks&2& $757.4 (1.2)-i\,80.8 (0.9)$&13.2(1.6)&9.1(1.3) &18.5(1.1) \\
        \hline
        Protopopescu&2& $757.3(1.2)-i\,81.0(0.8)$&10.6(1.5)&8.3(1.3)&9.5(1.0) \\
        \hline
        Estabrooks&3& $766.7(1.5)-i\,45.0(0.8)$&29.0(2.3)&26.8(2.0)&3.2(1.7) \\
        \hline
        Protopopescu&3&$764.8(1.2)-i\,44.4(1.0)$&34.2(2.3)&35.1(2.3)&9.5(1.8) \\
        \hline
        All&3& $766.1(1.8)-i\,44.1(0.9)$&32.9(2.2)&32.3(2.0)&8.8(1.3) \\
        \hline
    \end{tabular}
\end{table*}
%%%%%%%%%%%%%%
%%%%%%%%%%%%%%

The training data is given to a neural network with exactly the same details as the neural network used in the toy example of Sec.~\ref{subsec:ToyExample}. For each case, the neural networks are run $N=100$ times. The value of 100 is chosen somewhat arbitrarily, however, we performed numerical tests described in App.~\ref{App:Convergence} indicating that the sample size is enough for convergence. Randomness in optimization such as the batch selection leads to different final results. These $N=100$ different results are used to calculate the uncertainty as described in Step 7. in Sec.~\ref{subsec:Methodology}.

A common concern to many applications of neural networks and AI in general is that a possible bias or prior in the pseudodata might lead to unnaturally accurate results. For example, even though the range of values in which the poles are retained is centered around the $\chi^2$ pole, it is possible that the pole values for randomly generated parameters that fall in this range tend to be clustered closer to the reference PDG results. To address this, we calculate the pole positions for each point of the training data ($10^6$) and take the average as provided in \cref{tab:PseudoComparison}. These positions can also be compared to the pole positions obtained previously for the two methods in Tab.~\ref{tab:chi2}. Quantitatively, this is encoded in the provided distances in \cref{tab:PseudoComparison} between the poles from the two methods and the average training data poles for each model and data set. For  $n=3$ and all data combinations, the average of the poles from the training data is quite far from the predictions of either method. The distance to either method is more than three times as large as the that between the methods themselves. This demonstrates that the particular choice of the training data cannot have been the reason for the better performance of the SBI method. For the $n=2$ cases, the average of the training data pole positions is closer to SBI result. This indicates that the data used to train the neural network might slightly influence the accuracy of the SBI method. However, the fact that the average training data pole is roughly equidistant from the predictions from the two methods in both cases indicates that this bias is not the primary cause of the success of the SBI method. Furthermore, we note that the distance from the SBI result to the reference values is much smaller than that to the training data, or between the $\chi^2$ result to the reference values as well, see \cref{fig:TemperatureRatios}.

%%%%%%%%%%%%%%%%%%%%%%%%%%%%%%%%%%%%%%%%%%
\section{Discussion}
\label{sec:Discussion}
%%%%%%%%%%%%%%%%%%%%%%%%%%%%%%%%%%%%%%%%%%

In the previous sections we have demonstrated that SBI can be more accurate than $\chi^2$ minimization in certain cases of model misspecification. In this section, we discuss how the neural network's predictions can be interpreted. To some extent, we are limited to speculation, because, in general neural networks are \emph{"black boxes"} which behavior cannot be easily inferred from the details of the network~\cite{grus2019data}.

We can however remark that this process can be understood as doing something related to Bayesian inference. In model fitting (whether or not misspecification is present) the desired result is a probability distribution for the fit parameters, $\vec{a}$. This distribution is conditional on the measured data $\vec{y}$. In other words, the distribution to be estimated is $P(\vec{a}|\vec{y})$. As mentioned in Sec.~\ref{subsec:Methodology}, the probability distribution used to generate the training parameters $\vec{a}_j$ can be interpreted as the prior probability distribution. Each point of training data is generated in proportion to the likelihood of values for a given $\vec{a}$, thus, if the set is large enough, it contains an implicit probability distribution for $P(\vec{y}|\vec{a})$.  Thus, the neural network has access to enough information to perform an approximate Bayesian inference. Due to the complex nature of neural networks it cannot be verified if this is exactly what is being done, but the success of the neural network supports this claim.

%%%%%%%%%%%%%%%%%%%%%%%%%%%%%%%%%%%%%%%%%%
\section{Conclusion}
\label{sec:Conclusion}
%%%%%%%%%%%%%%%%%%%%%%%%%%%%%%%%%%%%%%%%%%

We have presented a method of Simulation Based Inference and shown it to be 
a viable alternative method to the $\chi^2$ minimization in case of model misspecification both for a toy example and for the case of the actual $\pi\pi$ phase-shift data used to predict the $\rho(770)$-resonance parameter. In the latter case, the SBI, indeed provides resonance parameters closer to the reference values for two different models and two different data sets indicating the robustness of the method. Additionally, a classifier neural network was employed to determine which model is most likely to have generated both sets of data simultaneously. SBI was then employed to determine the best parameters for this model. These parameters can be applied, beyond determination of the $\rho(770)$ to study of three-body states such as the $a_1(1260)$. 

While it is difficult to exactly determine the reason why the SBI method outperforms the $\chi^2$ minimization method in the described cases of model misspecification, we have speculated that the effectiveness of the SBI is due to the additional information the method has access to. 

The SBI cannot outperform $\chi^2$ minimization in cases that are close to ideal, and for this reason, we do not believe that it will replace the latter method. However, the strength of the method presented here gives strong reason to believe that application to other situations where data (including possible ambiguities) can lead to model misspecification would allow for a more robust determination of pole parameters.

%%%%%%%%%%%%%%%%%%%%%%%%%%%%%%%%%%%%%%%%
\begin{acknowledgments}
We are grateful to Dany Lane, Saverio Perugini, Alex Tsai, Dave Ireland, Denny Lane Sombillo and Anastasios Belias for useful discussions regarding machine learning applications. We thank James Carlini, Christian Pichay, Antonio Iijima and Maria Voce for contributing to this research. We thank Michael Döring for helping in developing useful algorithms used in parts of this work. 
We also acknowledge the constructive suggestions made by the reviewers of the paper.
The work of MM was funded through the Heisenberg Programme by the Deutsche Forschungsgemeinschaft (DFG, German Research Foundation) – 532635001. 
\end{acknowledgments}
%%%%%%%%%%%%%%%%%%%%%%%%%%%%%%%%%%%%%%%%

%%%%%%%%%%%%%%%%%%%%%%%%%%%%%%%%%%%%%%%%%%%%%%%%%%%%%%%%%
\bigskip
%%%%%%%%%%%%%%%%%%%%%%%%%%%%%%%%%%%%%%%%%%%%%%%%%%%%%%%%%
%%%%%%%%%%%%%%%%%%%%%%%%%%%%%%%%%%%%%%%%%%%%%%%%%%%%%%%%%
\bibliography{main}
%%%%%%%%%%%%%%%%%%%%%%%%%%%%%%%%%%%%%%%%%%%%%%%%%%%%%%%%%

\clearpage
\begin{onecolumngrid}

\appendix

%%%%%%%%%%%%%%%%%%%%%%%%%%%%%%%%%%%%
\section{Additional Numerical Tests: Network Architecture}
\label{App:Architecture}
%%%%%%%%%%%%%%%%%%%%%%%%%%%%%%%%%%%%

% \cor{DS}{}{In this section, we perform several numerical tests regarding specific choices of the neural network.}

The architecture of the neural network is in general somewhat arbitrary. To provide some level of objectivity in the context of the current application, we perform a numerical test demonstrating that, while there is not one ideal choice, the architecture is at least credible.

The neural network employed in this paper (except for the ones used in this test)  chooses 4 hidden layers with 40, 30, 20 and 10 nodes respectively. To test whether the number of nodes is sufficient we perform the following procedure. We vary the total number of nodes in the neural network, starting by fixing the network at 4 layers with the first layer having 4 times the number of nodes in the final layer, the second having 3 times, and the third layer having 2 times.  We generate 10 sets of $M$ and $\Gamma$ with uniformly distributed values between 400-1200~MeV and 0-500~MeV, respectively. With these values, we generate corresponding sets of pseudodata using the Breit-Wigner-like function described in the toy model.  We then train the neural network 10 separate times. For each time, we calculate the Mean Squared Error (MSE) , which is the chosen loss function in this work, between the model's prediction and the parameters used for generation. The average MSE for each neural network is given in the left half of \cref{tab:architecture}. We find that, for a very small network, increasing the number of nodes significantly increases the accuracy of prediction, however, beyond a certain size, increasing the number of nodes has a vanishingly small effect. The neural network we work with in the main part, with 100 nodes, is close to the point at which increasing the size has only a small effect on the accuracy. Since increasing the size of the neural network comes with an increased computational cost, we choose to work with a neural network of 100 nodes.

The right side of \cref{tab:architecture} holds the number of nodes fixed at 100 and varies the number of layers. The nodes are distributed among the layers as closely as possible to the distribution in which each layer decreases by the number of nodes contained in the final hidden layer. We find that a neural network with only one layer is already reasonably accurate, but increasing the number of layers provides a modest improvement up to around 4 layers. Beyond this, increasing the number of layers starts to provide a sharp increase in the MSE.  

%%%%%%%%%%%%%%
%%%%%%%%%%%%%%
\begin{table}[h]
\caption{A numerical test about the ideal architecture of the neural network. The left two columns test the number of nodes used. The MSE referenced is the average mean squared error for 10 different parameters. In all rows in these columns the nodes are distributed in four layers, with nodes decreasing in integer multiples of the number of nodes in the final layer. The left two columns test the number of hidden layers used.  100 nodes is distributed among the given number of layers. In both sides of the figure, the neural network used in the main text of this paper with 100 nodes and four hidden layers is put in bold font.}
\label{tab:architecture}
\begin{tabular}{|c|c|}
\hline
\# of nodes & MSE \\
\hline
\hline
20   & 9444.05 \\
50  & 4737 \\
80 & 59.84 \\
\textbf{100}  & \textbf{53.5}\\
150  & 53.77  \\
500 & 53.2  \\
\hline
\end{tabular}
%%%
~~~~~~~~~~~~
%%%
\begin{tabular}{|c|c|}
\hline
\# of hidden layers & MSE \\
\hline
\hline
 1 & 61.05 \\
 2 &  61.42  \\
3 & 56.83 \\
 \textbf{4} & \textbf{53.5} \\
 5 & 62.35  \\
 9 & 28987 \\
\hline
\end{tabular}
\end{table}
%%%%%%%%%%%%%%
%%%%%%%%%%%%%%

%%%%%%%%%%%%%%%%%%%%%%%%%%%%%%%%%%%%
\section{Additional Numerical Tests: Convergence}
\label{App:Convergence}
%%%%%%%%%%%%%%%%%%%%%%%%%%%%%%%%%%%%

Here, we perform a numerical test to determine whether 100 samples of the neural network are sufficient and whether the result converges with the sample size. Our method is as follows: We use the data set employed in the first toy example in which two points have been intentionally altered. We generate a list of 2000 neural network predictions. This list is then randomly broken into samples of a given size, $N_s$. The average and standard deviation of each sample is computed. From these a quantity known as the Standard Error of the Standard Deviation (SESD) is computed. This quantity consists of taking the standard deviation of the list of the standard deviations for each set. The Mean of Standard Deviations (MSD) is also computed. To obtain a more reliable value, these quantities are averaged over 1000 random partitions of the list of data into samples of size $N_s$. The ratio of these two quantities gives an estimate of what fractional or percentage variation that might occur between the standard deviation of a single sample of size $N_s$ and the standard deviation of an arbitrarily large sample. These values are shown for $M$ and $\Gamma$ in the left two columns of Tab.~\ref{tab:SampleTest}. We find that the standard deviation of a typical sample of size $N_s=5$ might be off by 40\% but that reduces to 8-11\% for a sample of $N_s=100$ as is employed in this work.

Additionally, we compute the $\bar \chi^2_{\rm dof}$ by partitioning the list into samples of size $N_s$ as described in the preceding paragraph. The $\chi^2$ statistic is calculated using the average $\overline{x_i^s}$ $(x\in \{M,\Gamma\})$ and standard deviation $\sigma_{i}^s$ of sample $i\in\{1,2000/N_s\}$ as well as the average from the whole list excluding that sample $\overline{x_i}$ as
%%%%%%%%
\begin{align}
\bar \chi^2_{\rm dof}:=\sum_i (\overline{x_i^s}-\overline{x_i})^2/(\sigma_{i}^s/\sqrt{N_s})^2\,.
\end{align}
%%%%%%%%
Assuming that the list is large enough that differences between its average quantities and the average quantities in an infinite list are negligible, and assuming that error is distributed normally and thus the uncertainty is $\sigma_i^s/\sqrt{N_S}$, this statistic should follow the $\bar \chi^2$ distribution. If this distribution is divided by the number of degrees of freedom---in this case the number of samples in the list---it has an expected value of 1. Table \ref{tab:SampleTest} contains this measure for both $M$ and $\Gamma$ averaged over many random re-partitions of the list. It shows that for a sample of $N_s=5$, the sample size is not large enough to obtain a reasonable convergence and uncertainty may not be sufficient. However, with a sample of $N_s=100$, the size employed in this work, the $\bar\chi^2_{dof}$ approaches 1, indicating that this uncertainty can sufficiently account for uncertainties due to fluctuations in the neural network.

%%%%%%%%%%%%%%%%%%
%%%%%%%%%%%%%%%%%%
\begin{table}[h!]
\caption{The average $\bar \chi^2_{\rm dof}$ and average SESD/MSD ratio for $M$ and $\Gamma$  }
\label{tab:SampleTest}
\begin{tabular}{|c|c|c|c|}
\hline
$N_s$ & $\chi^2_{dof}$  & SESD/MSD for $M$ & SESD/MSD for $\Gamma$ \\
\hline
5 & 1.87 & 0.40 & 0.39\\
10 & 1.27 & 0.27 & 0.24 \\
25 & 1.08& 0.18 & 0.18 \\
50 & 1.04& 0.16 & 0.12 \\
100 & 1.03& 0.11 & 0.08\\
\hline
\end{tabular}
\end{table}
%%%%%%%%%%%%%%%%%%
%%%%%%%%%%%%%%%%%%

\end{onecolumngrid}

\end{document}